\documentstyle[12pt]{article}
\newcommand\be{\begin{equation}}
\newcommand\ee{\end{equation}}
\newcommand\ba{\begin{eqnarray}}
\newcommand\ea{\end{eqnarray}}

\newcommand{\cl}{\centerline}
\newcommand\bear{\begin{eqnarray*}}
\newcommand\eear{\end{eqnarray*}}

\begin{document}
\begin{titlepage}
\setlength{\textwidth}{5.0in}
\setlength{\textheight}{7.5in}
\setlength{\parskip}{0.0in}
\setlength{\baselineskip}{18.2pt}
\hfill
{\tt HD-THEP-03-4}
\begin{center}
{\large{\bf Lagrangean Approach to Gauge Symmetries for Mixed Constrained Systems and the Dirac Conjecture}}\par
\vskip 0.3cm
\end{center}
\begin{center}
{Heinz J. Rothe and Klaus D. Rothe}\par
\vskip 0.3cm
{Institut f\"ur Theoretische Physik}\par
{Universit\"at Heidelberg, Philosophenweg 16, D-69120 Heidelberg, Germany}
\footnote{email: h.rothe@thphys.uni-heidelberg.de\\
k.rothe@thphys.uni-heidelberg.de}
\cl{\today}
\end{center}

\begin{abstract}
\noindent
The gauge symmetries of a general dynamical system can be systematically obtained following either a Hamiltonean or a Lagrangean approach. In the
former case, these symmetries are generated, according to Dirac's
conjecture, by the first class constraints. In the latter approach
such local symmetries are reflected in the existence of so called gauge identities. The connection between the two becomes apparent, if one works with a first order Lagrangean formulation. 
We thereby confirm Dirac's conjecture. Our analysis applies to arbitrary
constrained systems with first and second class constraints, and thus extends
a previous analysis by one of the authors to such general systems.
We illustrate our general results in terms of several examples.

\bigskip\noindent
PACS: 11:10; 11:15; 11:30
\end{abstract}
\end{titlepage}

\section{Introduction}

The problem of revealing the gauge symmetries of a Lagrangean has 
been the subject of numerous investigations \cite{1}-\cite{7}. It has been conjectured 
by Dirac a long time ago \cite{8}, that the generators of these symmetries 
are the first class constraints in the Hamiltonean formulation, subject to certain restrictions on the gauge parameters. 

Local symmetries of a dynamical system have been studied both within the Lagrangean as well as the Hamiltonian framework. On the 
Lagrangean level there exists a well known algorithm \cite{9,10} for detecting the 
gauge symmetries of a Lagrangean. It has the merit of directly generating the 
transformation laws in configuration space, expressed in terms of an
independent set of arbitrarty functions, which leave the action invariant. Every one of the
gauge parameters parametrizing such a local symmetry 
is directly related to a so-called ``gauge identity".
The number of such parameters is equal to the number of independent gauge identities.

On the Hamiltonian 
level the relevant action whose vanishing variation leads to the Hamilton 
equations of motion, is the so called ``total Hamilton principal function". The transformation laws in phase space, which leave this action invariant, have
been conjectured by Dirac to be generated by the so-called first class constraints.
The number of such constraints is in general larger than the number of
gauge-identities of the Lagrangean formulation, referred to above. Hence restrictions must be imposed on the corresponding gauge parameters, in order to generate the gauge symmetries of the
theory \cite{2,BRR}. The demonstration of the equivalence of the two approaches for an arbitrary
dynamical system with mixed first and second class constraints is object of this paper, and serves to confirm Dirac's conjecture. The relation between the 
Lagrangean and Hamiltonean description is made apparent by working with
an equivalent (total) Lagrangean in the first order formulation, and is an
extension to mixed constrained systems of previous work by one of the authors
\cite{11}, where
the formalism was developed for the case of a purely first class
system involving only one primary constraint.

The organization of the paper is as follows. In section 2 we discuss the level by level generation of the constraints in the Hamiltonean approach,
and organize these constraints in the form of what we refer to as first and second class chains. This will allow us to restrict our attention in section 3
to effectively first class systems with a new Hamiltonean. The first class chains are then generated iteratively following
a Lagrangean algorithm. We show that with every first class
primary constraint there is associated a first class chain, terminating
in a gauge identity. We then establish a direct relation between these gauge
identities and the first class constraints as generators of the local 
symmetries, thus confirming Dirac's conjecture. We illustrate our results
in terms of two examples. One of them serves to confirm an assertion made
in this section . The other example has been quoted in the literature as a counter example to Dirac's conjecture, and we show in what sense the
Dirac conjecture holds.
We conclude in section 4, and leave two further instructive examples for the Appendices.

\section{Preliminaries}

Let us denote by $\Omega^{(0)}_{A}$ ($A = 1,2,\cdot\cdot\cdot$) the primary constraints associated with a second order Lagrangean $L(q,\dot q)$,
where $q_i,i=1,\cdots,n$ are coordinates in configuration space, and
let $H(q,p)$ be the corresponding canonical Hamiltonian evaluated on the 
primary surface. Following the Dirac algorithm, we generate {\it level by level}
the secondary constraints, by requiring level by level the conservation of 
the constraints in time
with respect to the total Hamiltonian
\be\label{HT}
H_T = H(q,p) + \sum_A \lambda_A \Omega^{(0)}_A\,.
\ee
In this way we can associate with each primary constraint 
$\Omega^{(0)}_A$ a chain of secondary constraints $\Omega^{(\ell)}_A$,
where the superscript denotes the level of the iterative procedure. 
Denote by $\Gamma^{(\ell)}$ the constrained surface defined by all
constraints generated up to level $\ell$. 
A new constraint at level $\ell + 1$ in the $A'th$ chain is generated if the Poisson bracket $\{\Omega^{(\ell)}_A,\Omega^{(0)}_B\}_{\Gamma^{(\ell)}}=0$ for all $B$, while $\{\Omega^{(\ell)}_A,H\}_{\Gamma^{(\ell)}}\neq 0$. 
In this case 
we define a new constraint $\Omega^{(\ell+1)}_A$ by
\be\label{constraint-iteration}
\Omega^{(\ell+1)}_A := \{\Omega^{(\ell)}_A,H(q,p)\}\,.
\ee
The algorithm comes to a halt if no new constraints are generated. For 
a particular A-chain $\Omega^{(\ell)}_A$ this can happen in two ways: 

i) For $\ell = N_A$, $\{\Omega^{(N_A)}_A,\Omega^{(0)}_B\}_{\Gamma^{(N_A)}} 
\neq 0$, for some $B$. In 
this case the requirement that $\Omega^{(N_A)}_A$ should vanish for all times 
leads to a restriction on the Lagrange multipliers. 

\noindent
ii) For some $N_A$, and all $B$,  $\{\Omega^{(N_A)}_A,\Omega^{(0)}_B\}_{\Gamma^{(N_A)}} 
= 0$, 
but also  $\{\Omega^{(N_A)}_A,H\}_{\Gamma^{(N_A)}} = 0$. In this case the chains ends 
in what we shall refer to as a ``gauge identity".  We call such chains {\it first class chains}, for reasons to become 
clear later. 

We are thus led to classify the primary constraints into those leading to
gauge identities, $\{\phi^{(0)}_a\}$, and those leading to restrictions on the Lagrange
multipliers, $\{\psi^{(0)}_\rho\}$. We shall denote the respective chains by $\phi^{(\ell)}_a$ $(\ell = 1,\cdots,N_a)$ and $\psi^{(\ell)}_\rho$ $(\ell = 1,\cdots,M_\rho)$.
In this way we arrive at the following table, with the final elements in a chain satisfying,
\[
\{\phi^{(N_a)}_a,H\}_{\Gamma^{(N_a)}} = 0\,,\quad 
\{\phi^{(N_a)}_a,\Omega^{(0)}_A\}_{\Gamma^{(N_a)}} = 0
\]
for all $A$, and
\footnote{Our level by level generation of the constraints, 
following the Dirac algorithm, presumes that a $\psi_\rho$-chain terminates 
once condition (\ref{psifinal}) is satisfied.}
\be\label{psifinal}
\{\psi^{(M_\rho)}_\rho,\Omega^{(0)}_A\}_{\Gamma^{(M_\rho)}} \ne 0\,,
\ee
for some $A$.

\be
\begin{array}{cccc|ccccc}
\phi^{(0)}_1&\phi^{(0)}_2&\cdots&\phi^{(0)}_N
&\psi^{(0)}_1&\psi^{(0)}_2&\cdots&\psi^{(0)}_M
\nonumber\\ 
\hline
\phi^{(1)}_1&\phi^{(1)}_2&\cdots&\phi^{(1)}_N
&\psi^{(1)}_1&\psi^{(1)}_2&\cdots&\psi^{(1)}_M
\nonumber\\
\cdot&\cdot&\cdots&\cdot&\cdot&\cdot&\cdots&\cdot
\nonumber\\
\cdot&\cdot&\cdots&\cdot&\cdot&\cdot&\cdots&\cdot
\nonumber\\
\cdot&\cdot&\cdots&\cdot&\psi^{(M_1)}_1&\cdot&\cdots&\cdot
\nonumber\\
\phi^{(N_1)}_1&\cdot&\cdots&\cdot&&\cdot&\cdots&\cdot
\nonumber\\
&\phi^{(N_2)}_2&\cdots&\cdot&&\psi^{(M_2)}_2&\cdots&\cdot
\nonumber\\
&&&\phi^{(N_N)}_N&&&&\psi^{(M_M)}_M
\end{array}
\nonumber
\ee 


We next show that the primary constraints $\phi^{(0)}_a=0$ leading to gauge identities are necessarily first class, while the primary constraints
$\psi^{(0)}_\rho=0$ are all second class. To this end we make use
of the following

\bigskip\noindent
{\bf Assertion}

\bigskip

For any two given chains generated level by level from $\Omega^{(0)}_A$
and  $\Omega^{(0)}_B$, the following relations hold,
\be\label{crossrelations}
\{\Omega^{(0)}_A,\Omega^{(K_B)}_B\}_{\Gamma^{(K_B)}} = -\{\Omega^{(1)}_A,\Omega^{(K_B-1)}_B\}_{\Gamma^{(k_B)}} 
\cdots = (-)^r\{\Omega^{(r)}_A,\Omega^{(K_B-r)}_B\}_{\Gamma^{(K_B)}}
\ee
where $K_B$ labels the terminating element of the $B$-chain.

In order to prove the assertion made above, we make repeated use of 
\bear
\{\Omega^{(\ell)}_A,\Omega^{(\ell')}_B\} &=& 
\{\Omega^{(\ell)}_A,\{\Omega^{(\ell'-1)}_B,H\}\}\\
&=&-\{\{\Omega^{(\ell)}_A,H\},\Omega^{(\ell'-1)}_B\}
+ \{\{\Omega^{(\ell)}_A,\Omega^{(\ell'-1)}_B\},H\}
\eear
or
\be\label{PB1}
\{\Omega^{(\ell)}_A,\Omega^{(\ell')}_B\}= -\{\Omega^{(\ell+1)}_A,\Omega^{(\ell'-1)}_B\}
+ \{\{\Omega^{(\ell)}_A,\Omega^{(\ell'-1)}_B\},H\}
\ee
where we have made use of the Jacobi identity and of 
(\ref{constraint-iteration}). We begin with
\bear
\{\Omega^{(0)}_A,\Omega^{(K_B)}_B\} &=&-\{\Omega^{(1)}_A,\Omega^{(K_B-1)}_B\} +\{\{\Omega^{(0)}_A,\Omega^{(K_B-1)}_B\},H\}\\
&=&-\{\Omega^{(1)}_A,\Omega^{(K_B-1)}_B\} + \{\Phi^{(K_B-1)},H\}\,,
\eear
where $\Phi^{(K)}$ stands for a linear combination of the constraints
up to level $K$.
Hence
\be\label{PB2}
\{\Omega^{(0)}_A,\Omega^{(K_B)}_B\} =-\{\Omega^{(1)}_A,\Omega^{(K_B-1)}_B\} + \Phi^{(K_B)}\,.
\ee
 
The second step of the reduction will require twice the use of the
Jacobi identity.
Proceeding as above, we have from (\ref{PB1})
\be\label{PB3}
\{\Omega^{(1)}_A,\Omega^{(K_B-1)}_B\} = -\{\Omega^{(2)}_A,\Omega^{(K_B-2)}_B\} + \{\{\Omega^{(1)}_A,\Omega^{(K_B-2)}_B\},H\}\,.
\ee
For the evaluation of the second term on the r.h.s. we observe that, making
again use of (\ref{constraint-iteration}) and the Jacobi identity, we have
\ba\label{PB4}
\{\Omega^{(1)}_A,\Omega^{(K_B-2)}_B\} &=& \{\{\Omega^{(0)}_A,H\},\Omega^{(K_B-2)}_B\} \nonumber\\
&=& -\{\Omega^{(0)}_A,\Omega^{(K_B-1)}_B\} + \{\Phi^{(K_B-2)},H\}\nonumber\\
&=& \Phi^{(K_B-1)}\,.
\ea
Substituting this expression into (\ref{PB3}) we conclude that
\[
\{\Omega^{(1)}_A,\Omega^{(K_B-1)}_B\}_{\Gamma^{(K_B)}} 
= -\{\Omega^{(2)}_A,\Omega^{(K_B-2)}_B\}_{\Gamma^{(K_B)}}\,.
\]
Proceeding in this way, the above assertion (\ref{crossrelations}) follows.
We now prove the following corollar:

\bigskip\noindent
{\it Corollar}: {\it All first class primaries lead to gauge identities.} 

From the level-by-level iterative construction
of the constraints it follows that the Poisson brackets of all $\phi^{(0)}_a$ with the elements of the $\phi$-chains vanish weakly, since they end in gauge identities.
Furthermore, our level by level procedure implies that $\{\phi^{(0)}_a,\psi^{(\ell)}_\rho\} \approx 0$ for $\ell < M_\rho$ and all $a$.\footnote{Here and in the following, 
$\{\Omega^{(0)}_A,\Omega^{(\ell)}_B\} \approx 0$ means that the Poisson bracket 
vanishes on $\Gamma^{(\ell)}$.} To show that this is also true for  $\ell = M_\rho$, we distinguish two cases:

i) The length of the $\phi_a$-chain is larger or equal to that of the $\psi_\rho$-chain, i.e.
$N_a \geq M_\rho$.  From (\ref{crossrelations}), we conclude that
$\{\phi^{(0)}_a,\psi^{(M_\rho)}_\rho\} \approx 
(-)^{M_\rho}\{\phi^{(M_\rho)}_a,\psi^{(0)}_\rho\}$, which vanishes weakly
since $M_\rho \leq N_a$.
This result holds for all pairs for which $N_a \geq M_\rho$.

ii) The $\phi_a$-chain is shorter than the $\psi_\rho$-chain, i.e.
$N_a < M_\rho$;  Assume $\phi^{(0)}_a$ is first class, but does not 
lead to a gauge identity. Since $\{\phi^{(N_a)}_a,\phi^{(0)}_b\}\approx 0$
for all $b$, it then follows that for some $\rho$, 
$\{\phi^{(N_a)}_a,\psi^{(0)}_\rho\} \not\approx 0$. But according to 
(\ref{crossrelations}) 
this implies that $\{\psi^{(N_a)}_\rho,\phi^{(0)}_a\} \not\approx 0$, which 
contradicts the assumed first class character of $\phi^{(0)}_a$. Hence
$\phi^{(0)}_a$ must lead to a gauge identity.

This proves our corrolar.
We now show that the elements of the $\psi$-chains are second class
\footnote{We assume the $\psi$-chains to form an irreducible set in the
sense that all primaries leading to gauge identities have been isolated.}
. Define the $M_\rho \times M_\sigma$ matrix $Q$ with elements  
\be\label{Q}
Q_{\rho\sigma} := \{\psi^{(0)}_\rho,\psi^{(M_\sigma)}_\sigma\}\,.
\ee
Note that $Q$ is not an antisymmetric matrix.
Making again use of our assertion (\ref{crossrelations}),
we deduce the following properties of the  $\psi$-chains
\footnote{A similar reasoning and terminology has been used in ref. \cite{Shirzad}, where a chain-by-chain generation of
constraints has been considered. For our purposes, a level-by-level procedure
is important in order to generate the gauge identities.}
.

i) If $Q_{\rho\rho} \neq 0$ and $Q_{\rho\sigma} = 0$ for $\rho \neq \sigma$,  we obtain a ``self-conjugate $\psi$-chain".
This chain will involve an even number of elements. Indeed, if the
number of elements were odd, our reduction formula 
(\ref{crossrelations}) would lead in the 
final step to $\{\psi^{(\frac{M_\rho}{2})}_\rho,\psi^{(\frac{M_\rho}{2})}_\rho\}$,
which vanishes, in contradiction to our initial assumption that 
$\{\psi^{(M_\rho)}_\rho,\psi^{(0)}_\rho\} \not\approx 0$.

ii) If $Q_{\rho\sigma}\neq 0$ for $\rho\neq \sigma$, and $Q_{\rho\rho} = Q_{\sigma\sigma} = 0$, we obtain a pair of ``cross-conjugate" chains 
$(\psi_\rho,\psi_\sigma)$ of equal length. Indeed, suppose their length is unequal. Consider 
the shorter chain labelled by $\psi_\rho$, whose final element
$\psi^{(M_\rho)}_\rho$ has a non-vanishing Poisson bracket with
$\psi^{(0)}_\sigma$. Following the reduction procedure discussed above we would
conclude that $\{\psi^{(0)}_\rho,\psi^{(M_\rho)}_\sigma\} \not\approx 0$ for
$M_\rho <  M_\sigma$,  in contradiction to the algorithm generating the constraints.

iii) If $Q_{\rho\rho}$ and/or $Q_{\sigma\sigma}$, as well as $Q_{\rho\sigma}$ ($\rho \neq \sigma$) are non-vanishing we have ``mixed conjugate
chains", and joining the arguments of i) and ii) we conclude that 
such mutually linked chains are all of equal length and involve each
an even number of constraints.

From here we conclude that the $\psi$-chains form a second-class system. We assume them to form an irreducible set with
\be\label{detQ}
\det Q \not\approx 0\,.
\ee

Consider now the total Hamiltonian (\ref{HT}). In the new notation $\Omega^{(0)}_A = (\phi^{(0)}_a,\psi^{(0)}_\rho)$ it reads,
\[
H_T = H + \sum_a \lambda_a \phi^{(0)}_a
+  \sum_\rho \xi_\rho \psi^{(0)}_\rho \,.
\]
The persistence condition $\{\psi^{(M_\rho)}_\rho,H_T\}\approx 0$ together with  (\ref{detQ}) implies the fixation of all Lagrange multipliers associated with the $\psi^{(0)}_\rho$:
\[
\xi_\rho = - \sum_\sigma Q^{-1}_{\rho\sigma}\{\psi^{(M_\sigma)}_\sigma,H\}\,.
\]
Implementing these $\xi_\rho$, we are are led to define a new total Hamiltonian
\[
H^\star_T = H^\star + \sum_a \lambda_a \phi^{(0)}_a\,,
\]
where
\be\label{Hstar}
H^\star = H - \sum_\rho\psi^{(0)}_\rho Q^{-1}_{\rho\sigma}\{\psi^{(M_\sigma)}_\sigma,H\}
\ee
has weakly vanishing Poisson brackets with all the constraints, and therefore is first class
\footnote{Note that because of (\ref{constraint-iteration}), $H$ has
weakly vanishing Poisson brackets with all the constraints except for the
last members of the $\psi$-chains.}
.
We now repeat the level by level generation of the constraints in terms of
$H^\star$. We have
\[
\phi^{\star(\ell+1)}_a := \{\phi^{\star(\ell)}_a,H^\star\}
 \approx \phi^{(\ell+1)}_a
\]
and
\[
\psi^{\star(\ell+1)}_\rho := \{\psi^{\star(\ell)}_\rho,H^\star\}
 \approx \psi^{(\ell+1)}_\rho
\]
where $\approx$ means evaluation on the
constraint surface defined by all constraints up to
level $\ell$. Hence the ``stared" constraints are just linear combinations
of the constraints generated by $H_T$.

Since the $\phi^{(0)}_a$ and $H^\star$ are first class,
it follows by a well know theorem, that $\phi^{\star(1)}_a$ is first class, and 
proceeding iteratively in this way, we conclude that all $\phi^{\star(\ell)}_a$
are first class. 
From here on we drop the $\star$ on the iteratively generated constraints via $H^\star$, and only keep the $\star$ on $H^\star$ and $H^\star_T$, in order to remind the reader of this fact.

\section{Gauge identities and Dirac's conjecture}

Following our notation of the previous section, 
let $\phi^{(0)}_{a}$ ($a = 1,2,\cdot\cdot\cdot,N$) and
$\psi^{(0)}_{\sigma}$ ($\sigma = 1,2,\cdot\cdot\cdot,M$) be 
respectively the first and
second class primary constraints associated with a second order Lagrangean $L(q,\dot q)$,
where $q_i,i=1,\cdots,n$ are coordinates in configuration space, and
let $H^\star (q,p)$ be the Hamiltonian defined in (\ref{Hstar}). We assume these constraints to have been organized as
discussed in the previous section. 

One readily verifies that the Euler-Lagrange equations 
associated with the first order (total) Lagrangean
\footnote{The following procedure is 
a generlization of the formalism developed in \cite{11}. When refering to that reference, the reader should be aware of a number of notational differences.}
\be\label{Ltotal} 
L_T(q,p,\dot q,\dot p,\lambda,\dot\lambda) = \sum^n_{i=1}p_i\dot q_i 
- H^\star(q,p) - \sum_{a=1}^N\lambda^{a}\phi^{(0)}_{a}\,,
\ee
suplemented with the second class primary constraints
$\psi^{(0)}_\sigma = 0$,
reproduces the Hamilton equations of motion including the primary constraints,
if we regard $q_i,p_i$ and $\lambda^{a}$ as coordinates in an 
$2n+N$ dimensional configuration space.
We now write (\ref{Ltotal})
in the form 
\be\label{Ltotal-general}
L_T = \sum^{2n+N}_{\alpha=1} a_\alpha(Q)\dot Q_\alpha - H^\star_T(Q)\,,
\ee
where
\be\label{general-coordinates}
Q_\alpha := (\vec q,\vec p,\lambda^1,\cdot\cdot\cdot,\lambda^{N})
\ee
and
\be\label{Htotal}
H^\star_T(Q) = H^\star(Q_1,\cdot\cdot\cdot,Q_{2n}) + \sum_{a=1}^N 
Q_{2n+a}\phi^{(0)}_{a}\,.
\ee
The non-vanishing elements of $a_{\alpha}$ are given by $a_{i}=Q_{n+i}=p_i$ 
($i=1\cdot\cdot\cdot,n$).
The $2n+N$ components of the Euler derivative are given by
\footnote{Our definition of the Euler derivative differs
from that of ref. \cite{11} by a minus sign.}
\ba\label{Euler-derivative}
E^{(0)}_\alpha &=& 
\frac{d}{dt}\left(\frac{\partial L_T}{\partial\dot Q_\alpha}\right)
-\frac{\partial L_T}{\partial Q_\alpha}\\
&=& -\sum^{2n+N}_{\beta=1}F^{(0)}_{\alpha\beta}\dot Q_\beta
+ K^{(0)}_\alpha\,,
\ea
 with
\be\label{Falphabeta}
F^{(0)}_{\alpha\beta}= \partial_\alpha a_\beta - \partial_\beta a_\alpha \,.
\ee
$F^{(0)}$ is the $(2n+N)\times(2n+N)$ matrix
\be
{\bf F}^{(0)} = \left(
\begin{array}{ccc}
{\bf 0}&{\bf -1}&\vec 0\cdots \vec 0\\
{\bf 1}&{\bf 0}&\vec 0\cdots \vec 0\\
{\vec 0}^T&{\vec 0}^T&0\cdots 0\\
\cdot&\cdot&\cdot\\
\cdot&\cdot&\cdot\\
\cdot&\cdot&\cdot\\
{\vec 0}^T&{\vec 0}^T&0\cdots 0\\
\end{array}\right)\,,
\ee
and  
\be
K^{(0)}_\alpha = \frac{\partial H^\star_T}{\partial Q_\alpha}\,,
\ee
where ${\bf 1}$ is a $n\times n$ unit martix, $\vec 0$ are $N$-component Null column vectors
(associated with the absence of $\dot\lambda^a$ in $L_T$), and $\vec 0^T$ is the transpose of $\vec 0$.

The variation of the total action
\be\label{Stotal} 
S_T = \int dt\ L_T(Q,\dot Q)
\ee
is given by
\be\label{delta-Stotal}
\delta S_T = -\sum_\alpha\int dt\ E^{(0)}_\alpha\delta Q_\alpha\,,
\ee
where we have dropped a boundary term.
The left-zero modes of ${\bf F}^{(0)}$ are given by
\be
\vec v^{(0)}(a) = \left(\vec 0,\vec 0,{\hat n}(a)\right)\,,
\ee
where ${\hat n}(a)$ is a $N$-component unit vector   
with the only non-vanishing component in the $a$'th  
place. Hence we recover the (first class) primary constraints:
\be\label{vdotE0}
\vec v^{(0)}(a)\cdot\vec E^{(0)} = \vec v^{(0)}(a)\cdot\vec K^{(0)} = \phi^{(0)}_{a}\,.
\ee

We now adjoin the time derivative of the primary constraints 
to $\vec E^{(0)}$ and construct the $2n+2N$ component (level one) vector $\vec E^{(1)}$:
\be\label{E1}
{\vec E}^{(1)} = \left(
\begin{array}{c}
\vec E^{(0)}\\
\frac{d}{dt}{\vec\phi}^{(0)}\\
\end{array}\right)\,,
\ee
where $\vec\phi^{(0)} = (\phi^{(0)}_1,\cdot\cdot\cdot,\phi^{(0)}_{N})$. By construction $\vec E^{(1)}$ vanishes on shell, i.e. for 
$\vec E^{(0)} = 0$. 
The components of ${\vec E}^{(1)}$, which we label by $\alpha_1$, can be 
written in the form
\be
E^{(1)}_{\alpha_1} = -\sum_{\alpha}F^{(1)}_{\alpha_1\alpha}\dot Q_{\alpha} 
+K^{(1)}_{\alpha_1}(Q)\,,
\ee
where 
\be
{\vec K}^{(1)} = \left(
\begin{array}{c}
\vec K^{(0)}\\
\vec 0
\end{array}\right)\,,
\ee
and ${\bf F^{(1)}}$ is now the {\it rectangular} matrix 
\be\label{F1}
{\bf F}^{(1)} = \left(
\begin{array}{ccc}
{\bf 0}&{\bf -1}&\vec 0\cdots\vec 0\\
{\bf 1}&{\bf 0}&\vec 0\cdots\vec 0\\
{\vec 0}^T&{\vec 0}^T&0\cdots 0\\
\cdot&\cdot&\cdot\\
\cdot&\cdot&\cdot\\
\cdot&\cdot&\cdot\\
{\vec 0}^T&{\vec 0}^T&0\cdots 0\\
{-\nabla}\phi^{(0)}_1&{-\tilde\nabla}\phi^{(0)}_1&0\cdots 0\\
\cdot&\cdot&\cdot\\
\cdot&\cdot&\cdot\\
\cdot&\cdot&\cdot\\
{-\nabla}\phi^{(0)}_{N}&{-\tilde\nabla}\phi^{(0)}_{N}&0\cdots 0\\
\end{array}\right)
\ee
Here
\be\label{nabla}
{\nabla} := (\partial_{1},\cdot\cdot\cdot,\partial_{n})\,,\quad
{\tilde\nabla} := (\partial_{n+1},\cdot\cdot\cdot,\partial_{2n})\,.
\ee
We seek new constraints by looking for left zero modes of ${\bf F^{(1)}}$. 
They are $N$ in number, and are given by
\be\label{zeromode1}
\vec v^{(1)}(a) := \left(-{\tilde\nabla}\phi^{(0)}_{a},{\nabla}\phi^{(0)}_{a},
\vec 0,\hat e^{(0)}(a)\right)\,,
\ee
where $\hat e^{(0)}(a)$ is an $N$-component unit vector, with the only non-vanishing component in the $a$'th position. This leads to
\ba\label{level-one}
\vec v^{(1)}(a)\cdot\vec E^{(1)}= \vec v^{(1)}(a)\cdot\vec K^{(1)} 
&=& \frac{\partial\phi^{(0)}_{a}}{\partial q_i}
\frac{\partial H^\star_T}{\partial p_i}
-\frac{\partial H^\star_T}{\partial q_i}
\frac{\partial\phi^{(0)}_{a}}{\partial p_i}\nonumber\\
&=& \{\phi^{(0)}_{a},H^\star_T\}\,,
\ea
or
\be\label{vdotE1}
\vec v^{(1)}(a)\cdot\vec E^{(1)} = \{\phi^{(0)}_{a},H^\star\}  
-\sum_{c} \lambda^{c} \{\phi^{(0)}_{c},\phi^{(0)}_{a}\}\,,
\ee
which by constuction vanish on shell (i.e., for $\vec E^{(0)} = \vec 0$). Poisson brackets will always be understood
to be taken with respect to the canonically conjugate variables $q_i$ 
and $p_i$. 
For the purpose of illustration we suppose that the second term on the right hand side vanishes on the surface defined by the primary constraints,
and that we have a new constraint $\phi^{(1)}_a = 0$, with
\footnote{If this is not the case, the algorithm stops.}
\be\label{constraint1}
\phi^{(1)}_a := \{\phi^{(0)}_a,H^\star\}\,,
\ee
which is only a function of $q$ and $p$.
Hence from (\ref{constraint1}) and (\ref{vdotE1}),
\be 
\phi^{(1)}_{a} = \vec v^{(1)}(a)\cdot\vec E^{(1)}   
+\sum_{c} \lambda^{c} \{\phi^{(0)}_{c},\phi^{(0)}_{a}\} \,,
\ee
or
\be\label{Phi1}
\phi^{(1)}_{a} = \vec v^{(1)}(a)\cdot\vec E^{(1)} +  
\sum_{b,c} \lambda^{c}C^{[000]}_{cab}
(\vec v^{(0)}(b)\cdot\vec E^{(0)})\,,
\ee
where the coefficients $C^{[000]}_{cab}$ are defined by 
\be  
\{\phi^{(0)}_{c},\phi^{(0)}_{a}\} = 
\sum_{c}C^{[000]}_{cab}\phi^{(0)}_{b}\,,
\ee 
and use has been made of (\ref{vdotE0}). Note that $\phi^{(1)}_a$ is again a function of only $q$ and $p$.
  
We now repeat the process and adjoin the time derivative of 
the constraint (\ref{Phi1}) to the equations of motion to construct $\vec E^{(2)}$:
\be
{\vec E}^{(2)} = \left(
\begin{array}{c}
\vec E^{(0)}\\
\frac{d}{dt}\vec\phi^{(0)}\\
\frac{d}{dt}\vec\phi^{(1)}
\end{array}\right)\,,
\ee
where $\vec\phi^{(0)}$ and $\vec\phi^{(1)}$ are $N$  
component column vectors. This leads to a matrix ${\bf F}^{(2)}$. As we continue with this iterative process, the number of zero modes generated at each new level will in general be reduced, as ``gauge identities" are being generated along the way
(see below). Hence the number of components  of $\vec\phi^{(\ell)}$ will
in general decrease, as the level $\ell$ increases.

The constraints $\phi^{(\ell)}_{a}$ with $\ell\ge 1$ can be iteratively 
constructed from the recursion relation, 
\be\label{identity}
\phi^{(\ell+1)}_{a} = 
\vec v^{(\ell+1)}(a)\cdot\vec E^{(\ell+1)} 
+\sum^{\ell}_{\ell'=0}\sum_{b,c}\lambda^{c}
C^{[0\ell\ell']}_{cab}
\phi^{(\ell')}_{b} \,, \quad  \ell \ge 0
\ee
where,
\be
\phi^{(\ell+1)}_{a} := \{\phi^{(\ell)}_{a},H^\star\}
\ee
and the sum over $b$ in (\ref{identity}) runs over all constraints $\phi^{(\ell')}_b$ at level $\ell'$,
\footnote{Note that this set of constraints may be smaller in number than the set of (first class) primary constraints.}
The coefficients $C^{[0\ell\ell']}_{cab}$ are structure functions defined by 
\be\label{structure-constants}  
\{\phi^{(0)}_{c},\phi^{(\ell)}_{a}\} = 
\sum^\ell_{\ell'=0}\sum_{b}C^{[0\ell\ell']}_{cab}
\phi^{(\ell')}_{b}\,.
\ee
The zero modes $\vec v^{(\ell)}(b)$ have the 
following generic form :
\be\label{generic-eigenvectors}
\vec v^{(\ell+1)}(b) = (-\tilde\nabla\phi^{(\ell)}_{b},\nabla\phi^{(\ell)}_{b},\vec 0,
\hat e^{(\ell)}(b)) \,,\quad  \ell \ge 0 \,.
\ee
Here  $\hat e^{(\ell)}(b)$ is a unit vector with the only non-vanishing component at the position of the 
constraint $\phi^{(\ell)}_{b}$ in the array $(\vec\phi^{(0)}, 
\vec\phi^{(1)},\cdot\cdot\cdot,\vec\phi^{(\ell)})$ appearing 
in the expression for $\vec E^{(\ell)}$.
The entire iterative process will come to a halt at level $\ell = N_a +1$, when
\be\label{final-element}
\phi^{(N_a+1)}_a = \{\phi^{(N_a)}_a,H^\star\} = 
\sum_{\ell=0}^{N_a}\sum_{b}h^{[N_a\ell]}_{ab}\phi^{(\ell)}_{b}\,.
\ee
Making use of (\ref{final-element}) and (\ref{structure-constants}), and setting $\ell = N_a$ in (\ref{identity}), this equation takes the form
\be\label{gaugeidentity}
G_a := \vec v^{(N_a+1)}(a)\cdot\vec E^{(N_a+1)} -\sum_{\ell=0}^{N_a}\sum_b K^{[N_a\ell]}_{ab}\phi^{(\ell)}_b \equiv 0\,,
\ee
where
\be\label{K-coefficient}
K^{[N_a\ell]}_{ab}= h^{[N_a\ell]}_{ab}-\sum_{c}\lambda^cC^{[0N_a\ell]}_{cab}\,.
\ee
Eq. (\ref{gaugeidentity}) expresses the fact, that each
of the $\phi$-chains labelled by $a$ ends in a ``gauge identity" at the level $N_a+1$.

Iteration of (\ref{identity}), starting with $\ell=0$, allows us to express all the constraints in terms of scalar products $\vec v^{(\ell)}\cdot\vec E^{(\ell)}$.
Substituting the resulting expressions into (\ref{gaugeidentity}), and  
multiplying each of the gauge identities $G_{a}\equiv 0$ by an arbitrary function of time $\alpha_{a}(t)$, 
the content of all the identities can be summarized by an 
equation of the form
\be\label{general-identity}
\sum_{a=1}^N\sum_{\ell=0}^{N_a+1}\rho^{(\ell)}_{a}(Q,\alpha)
\left(\vec v^{(\ell)}(a)\cdot\vec E^{(\ell)}\right) \equiv 0 \,,
\ee
where
\be\label{El}
{\bf \vec E}^{(\ell)} = \left(
\begin{array}{c}
\vec E^{(0)}\\
\frac{d}{dt}\vec\phi^{(0)}\\
\frac{d}{dt}\vec\phi^{(1)}\\
\cdot\\
\cdot\\
\cdot\\
\frac{d}{dt}\vec\phi^{(\ell-1)}
\end{array}\right)\,.
\ee

Now, because of the generic structure of the eigenvectors (\ref{generic-eigenvectors}) we have from (\ref{El})
\be\label{vdotEl}
\vec v^{(\ell+1)}(a)\cdot \vec E^{(\ell+1)} = \sum^{2n}_{\alpha=1}
\left(v^{(\ell+1)}_\alpha(a) E^{(0)}_\alpha\right) 
+\frac{d\phi^{(\ell)}_{a}}{dt}
\,,\quad \ell = 0,\cdots,N_a
\ee
where $n$ is the number of coordinate degrees of freedom, and the constraints 
$\phi^{(\ell)}_a$ appearing on the RHS can be expressed, by iterating (\ref{identity}), in terms 
of scalar products $\vec v^{(k)}\cdot\vec E^{(k)}$, which in turn can be decomposed in the form (\ref{vdotEl}). Upon making 
a sufficient number of ``partial decompositions" $udv = d(uv) -vdu$, the identity (\ref{general-identity}) can be written in the form
\be\label{gaugeidentity2}
\sum_{a=1}^{N}\sum^{N_a}_{\ell=0}\epsilon^{(\ell)}_{a}\sum_{\alpha=1}^{2n+N}
\left(v^{(\ell+1)}_\alpha(a) E^{(0)}_\alpha\right)
+ \sum_{a=1}^N \tilde\epsilon_a \sum^{2n+N}_{\alpha=1}(v^{(0)}_\alpha(a)E^{(0)}_{\alpha})-\frac{dF}{dt}
\equiv 0\,,
\ee
where the 
$\{\epsilon^{(\ell)}_{a}\}$ and $\tilde\epsilon_a$ depend on the $N$ 
arbitrary functions of time $\{\alpha_{a}(t)\}$, as well as on the $Q_\alpha$'s 
and time derivatives thereof.  
This expression is of the form
\footnote{The minus sign has been introduced to cast the transformation 
laws in a standard form, when written in terms of Poisson brackets.}
\be\label{EdeltaQ}
-\sum_\alpha E^{(0)}_\alpha\delta Q_\alpha = \frac{dF}{dt}\,,
\ee
where
\be\label{deltaQ}
\delta Q_\alpha = -\sum^N_{a=1}\sum^{N_a}_{\ell=0} {\cal\epsilon}^{(\ell)}_{a}
v^{(\ell+1)}_\alpha (a)
- \sum_{a=1}^N \tilde\epsilon_a v^{(0)}_\alpha(a) \,.
\ee
For infinitessimal  ${\cal\epsilon}^{(\ell)}_{a}$ the time integral of the LHS is just the variation of the total action (\ref{delta-Stotal}).
Hence we conclude that the transformations (\ref{deltaQ}) 
leave the total action (\ref{Stotal}) invariant.
But because of the generic structure of the eigenvectors (\ref{generic-eigenvectors}) we have  from (\ref{deltaQ}),
\be\label{deltaq}
\delta q_i = \delta Q_i = \sum^N_{a=1}\sum^{N_a}_{\ell=0}{\cal\epsilon}^{(\ell)}_{a}
\frac{\partial\phi^{(\ell)}_{a}}{\partial p_i} 
= \sum^N_{a=1}\sum_{\ell=0}^{N_a}{\cal\epsilon}^{(\ell)}_{a}
\{q_i,\phi^{(\ell)}_{a}\}\,.
\ee
and
\be\label{deltap}
\delta p_i = \delta Q_{n+i} 
= - \sum^N_{a=1}\sum^{N_a}_{\ell=0}{\cal\epsilon}^{(\ell)}_{a}
\frac{\partial\phi^{(\ell)}_{a}}{\partial q_i} 
=\sum^N_{a=1}\sum_{\ell=0}^{N_a}{\cal\epsilon}^{(\ell)}_{a}
\{p_i,\phi^{(\ell)}_{a}\}\,,
\ee
while
\be\label{deltalambda1}
\delta\lambda^{a} = \delta Q_{2n+a} = -\tilde\epsilon_a \,.
\ee
Now, equation (\ref{identity}) can be written in the compact form 
\be
\vec v^{(\ell+1)}(a)\cdot \vec E^{(\ell+1)} = \{\phi^{(\ell)}_a,H_T^\star\}\,.
\ee
Hence it follows from (\ref{vdotEl}) that
\be
\sum_{\alpha}v^{(\ell+1)}_\alpha(a)E^{(0)}_\alpha = 
-\frac{d\phi^{(\ell)}_{a}}{dt}+
\{\phi^{(\ell)}_{a},H^\star_T\} \,,\quad  \ell = 0,\cdots,N_a \,.
\ee
Thus (\ref{gaugeidentity2}) can also be written in the form
\be\label{identity3}
\sum^N_{a=1}\sum_{\ell=0}^{N_a} \epsilon^{(\ell)}_{a}\left[-\frac{d\phi^{(\ell)}_{a}}{dt}+
\{\phi^{(\ell)}_{a},H^\star_T\} \right]-\sum_{a}
\delta\lambda^{a}\phi^{(0)}_{a} \equiv \frac{dF}{dt}\,,
\ee
where we have made use of (\ref{deltalambda1}).
The following argument suggests, that the function $F(q,p)$ is given by 
$F(q,p) = -\sum_{a=1}^N\sum_{\ell=0}^{N_a}{\cal\epsilon}^{(\ell)}_{a}
\phi^{(\ell)}_{a}(q,p)$.
To see this let us write out 
the LHS of (\ref{EdeltaQ}) in terms of the canonical variables $q$ and $p$ and Lagrange multipliers. From (\ref{Euler-derivative}) we see that
\be
\sum_\alpha E^{(0)}_\alpha\delta Q_\alpha = 
\sum_i\left[\left(\dot p_i+\frac{\partial H^\star_T}{\partial q_i}\right)
\delta q_i
- \left(\dot q_i-\frac{\partial H^\star_T}{\partial p_i}\right)\delta p_i\right]
+\sum_{a}\delta\lambda^{a}\phi^{(0)}_{a}\,.
\ee
Substituting the variations (\ref{deltaq}) and (\ref{deltap}) into this expression we arrive at
\be\label{EzerodeltaQ}
 -\sum_\alpha E^{(0)}_\alpha\delta Q_\alpha = 
\sum_{a=1}^{N}\sum^{N_a}_{\ell=0}
{\cal\epsilon}^{(\ell)}_{a}\left[-\frac{d\phi^{(\ell)}_{a}}{dt} 
+ \{\phi^{(\ell)}_{a},H_T^\star\}\right]  
-\sum_{a}\delta\lambda^{a}\phi^{(0)}_{a}\,.
\ee
This expression is of the form given by the LHS of (\ref{identity3}). If the $\delta Q_\alpha$ correspond to a symmetry of the total action (\ref{Stotal}), then it must be a total time derivative, and 
given by $\frac{dF}{dt}$. We now rewrite the RHS of (\ref{EzerodeltaQ}) in the form
\be
 \sum_{a=1}^{N}\sum_{\ell=0}^{N_a}
\left(\frac{d\epsilon^{(\ell)}_{a}}{dt}\phi^{(\ell)}_{a}+ 
\epsilon^{(\ell)}_{a}
\{\phi^{(\ell)}_{a},H_T^\star\}\right)  
-\sum_{a=1}^N\delta\lambda^{a}\phi^{(0)}_{a} 
-\frac{d}{dt}\left(\sum_{a=1}^N\sum_{\ell=0}^{N_a}{\cal\epsilon}^{(\ell)}_{a}
\phi^{(\ell)}_{a}\right)\,.
\ee
But quite generally
\be
\{\phi^{(\ell)}_{a},H_T^\star\} = \sum_{\ell'=0}^{\ell}\sum_{b}K^{[\ell\ell']}_{ab}\phi^{(\ell')}_{b}
\ee
where, because of (\ref{constraint-iteration}),
$K^{[\ell\ell']}_{ab}$ is given by
\be\label{Kcoefficient2} 
K^{[\ell\ell']}_{ab}=\delta_{ab}\delta_{\ell',\ell+1}\,,\quad \ell < N_a,
\ee
and by (\ref{K-coefficient}) for $\ell = N_a$.
Hence (\ref{identity3}) can also be written in the form
\be\label{recursion-identity}
\sum_{a=1}^N\sum_{\ell=0}^{N_a}\left({\dot{\cal\epsilon}}^{(\ell)}_{a}  
+\sum_{\ell',b}{\cal\epsilon}^{(\ell')}_{b}
K^{[\ell'\ell]}_{ba}\right)
\phi^{(\ell)}_{a} - \sum_{a=1}^N\delta\lambda^{a}\phi^{(0)}_{a} 
=  \frac{d\Phi}{dt}\,,
\ee
where
\be\label{surfaceterm}
\Phi(q,p) = F(q,p) +\left(\sum_{a=1}^N\sum_{\ell=0}^{N_a}{\cal\epsilon}^{(\ell)}_{a}
\phi^{(\ell)}_{a}(q,p)\right)\,.
\ee
Since the constraints are linearly independent, we conjecture that 
the $\{\epsilon^{(\ell)}_a\}$ are solutions to the following set of coupled differential equations
\be\label{recursion-relations}
{\dot\epsilon}^{(\ell)}_{a}  
+\sum_{\ell'}\sum_{b}{\cal\epsilon}^{(\ell')}_{b}K^{[\ell'\ell]}_{ba} = 0 
\,,\quad \ell\ge 1
\ee
and
\be\label{deltalambda}
\delta\lambda^{a} = {\dot{\cal\epsilon}}^{(0)}_{a} 
+ \sum_{\ell'}\sum_{b}{\cal\epsilon}^{(\ell')}_{b}K^{[\ell'0]}_{ba}\,, 
\ee
implying that 
\be\label{our-conjecture}
F(q,p) = -\sum_{a=1}^N\sum_{\ell=0}^{N_a}{\cal\epsilon}^{(\ell)}_{a}
\phi^{(\ell)}_{a}(q,p)
\ee
Equations (\ref{recursion-relations}) and (\ref{deltalambda}) are the familiar relations
obtained in the literature \cite{2,BRR}.

We now verify our conjecture in some examples. 

\subsection{Example 1}

In this section we apply the formalism to the case where the system 
exhibits two primary constraints $\phi^{(0)}_{a}, a = 1,2$, and one secondary constraint $\phi^{(1)}_2$, which in Dirac's language is generated 
from the presistency in time of $\phi^{(0)}_2$. The specific form of 
the Hamiltonian is irrelevant. {\it The constraints are assumed to be first class.}
For the case in question, our table in section 2 thus reduces to the form

\be
\begin{array}{c|c}
\phi^{(0)}_1&\phi^{(0)}_2\\ 
\hline
&\phi^{(1)}_2
\nonumber
\end{array}
\ee
In this case our conjecture can be checked generically.
 
The total Lagrangean $L_T$ is given by (\ref{Ltotal}) with $N=2$. Proceeding as in section 3, the primary 
constraints can be written in the form (\ref{vdotE0}), where $\vec v^{(0)}(1) 
= (\vec 0,\vec 0,1,0)$ and  $\vec v^{(0)}(2) = (\vec 0,\vec 0,0,1)$. 

Next we 
construct the vector (\ref{E1}), i.e. 
$\vec E^{(1)} = (\vec E^{(0)},\dot\phi^{(0)}_1,\dot\phi^{(0)}_2)$, and 
 the zero modes of the corresponding matrix ${\bf F}^{(1)}$, which are given by (\ref{zeromode1}), i.e., $\vec v^{(1)}(1) = (-\tilde\nabla\phi^{(0)}_1,\nabla\phi^{(0)}_1,0,0,1,0)$, 
$\vec v^{(1)}(2) = (-\tilde\nabla\phi^{(0)}_2,\nabla\phi^{(0)}_2,0,0,0,1)$.     Since 
only $\phi^{(0)}_2$ is assumed to lead to a new constraint, we are immediately left with one gauge identity at level 1, generated from $\phi^{(0)}_1$:
\be
G_1=\vec v^{(1)}(1)\cdot\vec E^{(1)} -\sum_{b=1}^2K^{[00]}_{1b}\phi^{(0)}_b
\equiv 0 \,,
\nonumber\\
\ee
or 
\be\label{GI1}
G_1=\vec v^{(1)}(1)\cdot\vec E^{(1)} -\sum_{b=1}^2K^{[00]}_{1b}\left(\vec v^{(0)}(b)\cdot\vec E^{(0)}\right) \equiv 0\,,
\ee   
which is nothing but (\ref{gaugeidentity}) with $N_1=0$. On the other hand $\phi^{(0)}_2$ gives rise to a new constraint $\phi^{(1)}_2$ at level 1, which is given by 
\be
\phi^{(1)}_2 = \vec v^{(1)}(2)\cdot\vec E^{(1)} + 
\sum^2_{b,c=1}\lambda^{c}C^{[000]}_{c2b}
\left(\vec v^{(0)}(b)\cdot\vec E^{(0)}\right)\,.
\ee
We are therefore led to construct 
\be
{\vec E}^{(2)} = \left(
\begin{array}{c}
\vec E^{(0)}\\
\frac{d}{dt}\phi^{(0)}_1\\
\frac{d}{dt}\phi^{(0)}_2\\
\frac{d}{dt}\phi^{(1)}_2
\end{array}\right)
\ee
and the corresponding rectangular matrix $F^{(2)}$ and its left zero modes, of 
which only the contraction of 
\be
\vec v^{(2)}(2) = (-\tilde\nabla\phi^{(1)}_2,\nabla\phi^{(1)}_2,0,0,0,0,1)
\ee
with $\vec E^{(2)}$ leads to a new equation, which is necessarily a 
gauge identity, since we have assumed that the system only possesses one 
secondary constraint $\phi^{(1)}_2$. The gauge identity at level 2 has the form
\be
G_2=\vec v^{(2)}(2)\cdot\vec E^{(2)}-\sum_{b=1}^2
K^{[10]}_{2b}\phi^{(0)}_{b}
- K^{[11]}_{22}\phi^{(1)}_2 \equiv 0 \,,\nonumber
\ee
or
\ba\label{GI2}
G_2=\vec v^{(2)}(2)\cdot\vec E^{(2)}&-&\sum^2_{b=1}K^{[10]}_{2b}
\left(\vec v^{(0)}({b})\cdot\vec E^{(0)}\right)\\
&-& K^{[11]}_{22}\left(\vec v^{(1)}(2)\cdot\vec E^{(1)}
+\sum_{b,c=1}^2\lambda^{c}C^{[000]}_{c2b}\left(\vec v^{(0)}(b)\cdot\vec E^{(0)}\right)\right) \equiv 0\,.\nonumber
\ea
Multiplying the gauge identities (\ref{GI1}) and (\ref{GI2}) by the arbitrary  
functions $\alpha_1(t)$ and $\alpha_2(t)$, respectively, and taking their sum, the resulting expression can be written in the form (\ref{general-identity}). Upon making a sufficient 
number of "partial differential decompositions", one finds, after some
algebra, that the information encoded in 
the gauge identities can be written in the form (\ref{gaugeidentity2}), 
where
\ba
\epsilon^{(0)}_1 &=& -\alpha_1\cr
\epsilon^{(1)}_2 &=& -\alpha_2\cr
\epsilon^{(0)}_2 &=& \dot\alpha_2 +\alpha_2K^{[11]}_{22}\cr
\delta\lambda^a &=&  
-\dot\alpha_2\sum_{c}\lambda^{c}C^{[000]}_{c2a} - 
\alpha_2K^{[10]}_{2a}
- \alpha_1 K^{[00]}_{1a}\cr
 &-& \alpha_2K^{[11]}_{22}\sum_{c}\lambda^{c}C^{[000]}_{c2a}
+\ddot\alpha_2\delta_{2a} - \dot\alpha_1\delta_{1a} 
+\frac{d}{dt}\left(\alpha_2K^{[11]}_{22}\right)\delta_{2a}\,.
\ea
The corresponding transformation laws for the the coordinates $q_i$ and 
momenta $p_i$ 
are given by (\ref{deltaq}) and (\ref{deltap}) with $\epsilon^{(1)}_1 = 0$.

Having obtained the the functions $\epsilon^{(0)}_{1},\epsilon^{(0)}_{2},
\epsilon^{(1)}_{2}$ and $\delta\lambda^a$ expressed 
in terms of $\alpha_1(t)$ and $\alpha_2(t)$, we can now 
check or conjecture that the $\epsilon^{(\ell)}_a$  are solutions to 
equations (\ref{recursion-relations}),
which in our case reduce to 
\[
\frac{d\epsilon^{(1)}_2}{dt} + \sum^2_{b=1}\epsilon^{(0)}_{b}
K^{[01]}_{b2}+\epsilon^{(1)}_2 K^{[11]}_{22} = 0\,,
\]
and that
\[
\delta\lambda^a = \dot\epsilon^{(0)}_a 
+ \sum_b \epsilon^{(0)}_b K^{[00]}_{ba} + \epsilon^{(1)}_2 K^{[10]}_{2a}\,,
\]
where we have made use of the fact that, because of the way in which the constraints
have been generated, $K^{[\ell1]}_{a1} = 0\ (a=1,2)$, $K^{[01]}_{22} = 1$,
$K^{[01]}_{12} = 0$, $C^{[001]}_{abc} = 0$, and
$K^{[00]}_{22} = - \sum_c \lambda^c C^{[000]}_{c22}$. One then verifies that the above equations are indeed satisfied.

A similar statement was shown to hold  for the example discussed in \cite{11}, where the the formalism was developed for a purely first class system involving only one primary constraint and an arbitrary number of secondary constraints.
\footnote{Reference \cite{11} contains some printing errors.
In particular the index ``$a$" in Eq. (56) of that reference takes the values
$a = 2,\cdots,M$, and $\alpha$ in Eq. (59) runs over $1,2,\cdots,2n+1$,
and not from $1$ to $7$, as stated in the paragraph following (59).}

\subsection{A counterexample to Dirac's conjecture?}

In the literature it has been stated that Dirac's conjecture is not always
correct. 
An example for this has been claimed \cite{12} to be given by the Lagrangean
\be\label{conjecture-Lagrangean}
L = \frac{1}{2}e^{q_2}{\dot q_1}^2\,.
\ee
In the following we analyze this system in detail, and point out some 
subtleties leading to an apparent clash with Dirac's conjecture.

The Lagrange equations of motion can be summarized by a single equation
\be\label{qdot}
\dot q_1 = 0\,.
\ee
Hence $q_2$ is an arbitrary function. 

Eq. (\ref{qdot}) does not possess a local symmetry. On the Hamiltonian level the system nevertheless 
exhibits two first class constraints, which, as we now show induce 
transformations of $q_1$ as well as of $q_2$ which are off-shell symmetries 
of the total action $\int dt\ L_T$, with $L_T$ defined in defined in (\ref{Ltotal}). However only one of the constraints generates a symmetry of the Hamilton equations of motion.  

From (\ref{conjecture-Lagrangean}) we obtain for the (only) primary constraint
\be\label{conjecture-primary}
\phi^{(0)} = p_2 = 0\,.
\ee
The canonical Hamiltonian evaluated on the primary surface is given by
\be
H(q,p) = \frac{1}{2}e^{-q_2}p_1^2\,,
\ee
and the first order total Lagrangean reads
\be
L_T(q,p,\lambda;\dot q,\dot p,\dot\lambda) =  \sum_i p_i\dot q_i-H(q,p)
-\lambda \phi^{(0)}\,.
\ee
Considered as a function of the "coordinates" $q_i$, $p_i$ and $\lambda$ and 
their derivatives, it has the form (\ref{Ltotal-general}), where $H^\star = H$,
$2n+N=5$, $Q_\alpha = (\vec q, \vec p,\lambda)$, 
and  $a_i = p_i$ ($i=1,2$) for the non-vanishing components of
$a_\alpha$. 

The Euler derivatives are given by (\ref{Euler-derivative}), where
$F^{(0)}_{\alpha\beta}=\partial_\alpha a_\beta - \partial_\beta a_\alpha$  
are the elements of a $5\times 5$ matrix with non-vanishing components
$F_{13} = F_{24} = -F_{31} = -F_{42} = -1$. 
The equations of motion are given by $E^{(0)}_\alpha = 0$, and 
yield the Hamilton equations of motion
\ba\label{Hamilton-equations}
\dot q_1 &-& e^{-q_2}p_1 = 0\cr
\dot q_2 &-& \lambda = 0\cr
\dot p_1 &=& 0\cr
\dot p_2 &-& \frac{1}{2}e^{-q_2}p_1^2 = 0\cr
p_2 &=& 0 \,.
\ea
We now proceed with the construction of the constraints and gauge identities 
as described in section 3. Since we have only one primary constraint, the 
formalism simplifies considerably. 

The matrix $F^{(0)}_{\alpha\beta}$ has one left zero-mode: 
\be
\vec v^{(0)} = (0,0,0,0,1)\,.
\ee
Its contraction with $\vec E^{(0)}$ just reproduces the primary constraint:
\be
\vec v^{(0)}\cdot\vec E^{(0)} = \phi^{(0)}\,.
\ee
Proceeding in the manner described in section 1, we construct $\vec E^{(1)}$ and corresponding left eigenvector of $F^{(1)}$, 
\[
v^{(1)} = (0,-1,0,0,0,1)\,,
\]
leading to the secondary constraint
\be
\vec v^{(1)}\cdot\vec E^{(1)} = \phi^{(1)}\,,
\ee
where
\be
\vec E^{(1)} = \left(
\begin{array}{r}
\vec E^{(0)}\\
\frac{d\phi^{(0)}}{dt}\\
\end{array}\right)
= \left(
\begin{array}{c}
\vec E^{(0)}\\
\frac{d}{dt}(\vec v^{(0)}\cdot\vec E^{(0)})\\
\end{array}\right)\,,
 \ee
and
\be
\phi^{(1)} = \frac{1}{2}e^{-q_2}p^2_1 \,.
\ee
Note that by construction
$\phi^{(1)} = 0$, on\ shell.
Since $\{\phi^{(0)},\phi^{(1)}\} = \phi^{(1)}$, $\phi^{(0)}$ and
$\phi^{(1)}$ form a first class system.
The algorithm is found to stop at level "2" where the following gauge identity 
is generated:
\be
\vec v^{(2)}\cdot\vec E^{(2)} + 
\lambda \vec v^{(1)}\cdot\vec E^{(1)} \equiv 0\,,
\ee
with
\be
\vec v^{(2)} = (-e^{-q_2}p_1,0,0,-\frac{1}{2}e^{-q_2}p^2_1,0,0,1)\,,
\ee
and  
\be
\vec E^{(2)} = \left(
\begin{array}{r}
\vec E^{(0)}\\
\frac{d\phi^{(0)}}{dt}\\
\frac{d\phi^{(1)}}{dt}
\end{array}\right)
= \left(
\begin{array}{c}
\vec E^{(0)}\\
\frac{d}{dt}(\vec v^{(0)}\cdot\vec E^{(0)})\\
\frac{d}{dt}(\vec v^{(1)}\cdot\vec E^{(1)})
\end{array}\right)\,.
\ee
The gauge identity can be reduced to the form 
\be
v^{(2)}_\alpha E^{(0)}_\alpha + \frac{d}{dt}(v^{(1)}_\alpha E^{(0)}_\alpha) 
+\frac{d^2}{dt^2}(v^{(0)}_\alpha E^{(0)}_\alpha) 
+\lambda[(v^{(1)}_\alpha E^{(0)}_\alpha) + 
\frac{d}{dt}(v^{(0)}_\alpha E^{(0)}_\alpha)] \equiv 0\,,
\ee
where a summation over $\alpha = 1,\cdot\cdot\cdot,5$ is understood.
Multiplying this expression by an arbitrary function of time $\epsilon(t)$, 
this identity becomes
\be\label{E0deltaQ}
\sum_\alpha E^{(0)}_\alpha\delta Q_\alpha = -\frac{dF}{dt}\,,
\ee
where
\be\label{deltaQalpha}
\delta Q_\alpha = \epsilon v^{(2)}_\alpha + (\lambda\epsilon-\dot\epsilon)v^{(1)}_\alpha + 
\left(\ddot \epsilon - \frac{d}{dt}(\lambda\epsilon)\right)v^{(0)}_\alpha\,.
\ee
In terms of the Hamiltonian variables, $q_i$, $p_i$ and $\lambda_i$, (\ref{deltaQalpha}) 
implies the following transformation laws
\ba\label{conjecture-deltaQ}
\delta q_1 &=& -\epsilon e^{-q_2}p_1 = \epsilon^{(1)}\{q_1,\phi^{(1)}\}\,,\cr
\delta q_2 &=& = \dot\epsilon - \lambda\epsilon =  \epsilon^{(0)}\{q_2,\phi^{(0)}\}\,,\cr
\delta p_1 &=& 0\,,\cr
\delta p_2 &=& -\frac{1}{2}\epsilon e^{-q_2}p_1^2 = \epsilon^{(1)}\{p_2,\phi^{(1)}\}\,,\cr
\delta \lambda &=& \ddot\epsilon - \frac{d}{dt}(\lambda\epsilon) 
= \dot\epsilon^{(0)}\,,
\ea
where
\be
\epsilon^{(0)} =\dot\epsilon - \lambda\epsilon\,,\quad
\epsilon^{(1)} = -\epsilon \,.
\ee
One readily verifies, that the $\epsilon^{(\ell)}$ satisfy the recursion relations (\ref{recursion-relations}), and that 
the function $F$ in (\ref{E0deltaQ}) is given by
\be
F = -\sum^2_{\ell=1}\epsilon^{(\ell)}\phi^{(\ell)}\,,
\ee
thus confirming once more our conjecture made in the previous section.  It is also worthwhile noting, that these recursion relations involve $\lambda$ itself, reflecting the fact, that the algebra of the first class constraints $\phi^{(0)}$ and $\phi^{(1)}$ only closes weakly.

We have thus verified that the transformations correspond to a symmetry of 
the action action (\ref{Stotal}). This symmetry is realized off shell, 
and requires the 
full set of transformation laws (\ref{conjecture-deltaQ}). In this sense the Dirac conjecture is verified also in this example. On the other hand, on the level of the 
Hamilton equations of motion we see that the constraints imply that 
$p_1 = p_2 = 0$ for all times. Hence the variations $\delta q_1$ and 
$\delta p_2$ do not generate a symmetry on the level of the equations of motion. Furthermore, $\delta\dot q_2 = \frac{d}{dt}\delta q_2 = \delta\lambda$ 
which is consistent with the Hamilton equations of motion (\ref{Hamilton-equations}). 

Let us summarize our findings. As we have shown, all the first class 
constraints take part in generating the symmetries of the total action. When interpreted in this way, Dirac's conjecture holds. On the other hand, 
not all of these constraints take necessarily part in generating symmetries of
the Hamilton equations of motion. In appendix A we present the corresponding analysis for the symmetries of the action associated with the quadratic Lagrangean (\ref{conjecture-Lagrangean}). There we
find that the corresponding action exhibits a local symmetry, whereas the equation of motion (\ref{qdot}) does not, in agreement with our findings in the
present section,
 
\section{Conclusion}
In this paper we have generalized the work of ref. \cite{11} to arbitrary 
Hamiltonian systems involving an arbitrary number of first and second class constraints. Using purely Lagrangean methods we have derived for
singular systems the 
local symmetries of the Hamiltonian equations of motion as well as
Hamilton principal function, 
and have discussed in which sense the Dirac conjecture holds. In particular we 
have presented an algorithm for generating the infinitessimal gauge 
transformations for systems involving first class constraints, and have shown that these have the form proposed by Dirac, i.e., 
$\delta q_i = \sum\epsilon^A\{q_i,\phi_A\}$, 
$\delta p_i = \sum\epsilon^A\{p_i,\phi_A\}$, where the $\{\phi^A\}'s$ are all the first class constraints of the theory. We have conjectured that the $\epsilon^A$'s 
are solutions to a coupled set of differential equations, in agreement 
with previous work \cite{2,BRR}. The general formalism developed has been applied to two simple systems, which illustrate in which sense the Dirac conjecture holds. One of these systems was argued in the literature to provide 
a counter example to the Dirac conjecture. 

\section*{Appendix A}

In this appendix we consider once more the Lagrangean (\ref{conjecture-Lagrangean}) and study the local symmetries of the corresponding action within the second order formalism.  While the equation of motion (\ref{qdot}) does not possess a local symmetry, the action does exhibit such a symmetry. The corresponding transformations of the coordinates are 
not obvious, but may be easily obtained using the standard Lagrangean algorithm for a second order Lagrangean. 
Connsider the Euler derivative (\ref{Euler-derivative}) with $\alpha \to i = 1,2$. 
It has the form
\be
E^{(0)}_i = \sum_j F^{(0)}_{ij}\ddot q_j + K^{(0)}_i\,,
\ee
where
\be
F^{(0)} = \left(
\begin{array}{cc}
e^{q_2}&0\\
0&0\\
\end{array}\right)
\ee
and
\be 
\vec K^{(0)} = \left(
\begin{array}{c}
e^{q_2}\dot q_2\dot q_1\\
-\frac{1}{2}e^{q_2}\dot q^2_1\\
\end{array}\right)
\ee
The left zero mode of $F^{(0)}$ is given by 
\be
\vec v^{(0)} = (0,1)\,.
\ee
Hence 
\be\label{varphi}
\vec v^{(0)}\cdot\vec E^{(0)} = \varphi\,,
\ee
where 
\be\label{varphi-constraint}
\varphi = -\frac{1}{2}e^{q_2}\dot q_1^2
\ee
is a constraint which depends only on the $q_i$'s and the velocities, and 
vanishes on shell. We now adjoin the time derivative of $\varphi$ to 
$\vec E^{(0)}$ and construct $\vec E^{(1)}$, as well as the matrix $F^{(1)}$, searching again for new left eigenvectors of  $F^{(1)}$. One finds 
\be\label{zeromode1}
\vec v^{(1)} = (\dot q_1,0,1)\,.
\ee
The algorithm stops at this point since 
\be
\vec v^{(1)}\cdot \vec E^{(1)} = -\dot q_2\varphi\,,
\ee
which implies the gauge identity
\be\label{gauge-identity4}
\vec v^{(1)}\cdot \vec E^{(1)} + \dot q_2(\vec v^{(0)}\cdot\vec E^{(0)}) 
\equiv 0\,.
\ee
Now
\be
 \vec v^{(1)}\cdot \vec E^{(1)} = \sum^2_{i=1} v^{(1)}_iE^{(0)}_i + \frac{d\varphi}{dt}\,.
\ee
Hence, upon making use of (\ref{varphi}), equation (\ref{gauge-identity4}) reduces to
\be
\sum_{i=1}^2\left[v^{(1)}_iE^{(0)}_i + (\frac{d}{dt}+\dot q_2)(v^{(0)}_iE^{(0)}_i)\right] \equiv 0\,.
\ee 
Multiplying this gauge identity by an arbitrary function of time $\epsilon(t)$ and making a partial differential decomposition, this identity takes the form
\be
\sum_iE^{(0)}_i\delta q_i + \frac{d}{dt}\left(\epsilon\varphi\right)\equiv 0\,,
\ee
where
\be
\delta q_i = \epsilon v^{(1)}_i + 
(\dot q_2\epsilon - \dot{\epsilon})v^{(0)}_i
\ee
or
\ba\label{deltaq12}
\delta q_1 &=& \dot q_1\epsilon\,,\cr
\delta q_2 &=& \dot q_2\epsilon - \dot{\epsilon}\,.
\ea
Consider now the variation $\delta L$. It is given by 
\be
\delta L(q,\dot q) = \frac{1}{2}e^{q_2}\dot q^2_1\delta q_2
+e^{q_2}\dot q_1\frac{d}{dt}\delta q_1\,. 
\ee
Upon making use of (\ref{deltaq12}) one then readily finds that
\be
\delta L = -\frac{d}{dt}(\epsilon\varphi)\,.
\ee
Hence the invariance of the action 
requires both, $\delta q_1$ and $\delta q_2$ to be different from zero. 
On the level of the equation of motion, on the other hand, $\delta q_1 = 0$, and $\delta q_2$ is an arbitrary function. Note that, in contrast to the 
Hamiltonian formulation, we have only one constraint in the Lagrangean approach, since there the concept of a primary constraint does not enter. The 
Lagrangean constraint $\varphi = 0$ is the analogue of the secondary 
constraint $\phi^{(1)}=0$ on the Hamiltonian level, which was shown not to 
generate a symmetry on 
the level of the equations of motion.

\section*{Appendix B} 

A simple example of a system with one primary and two secondary 
constraint is given by the following Lagrangean, considered in ref.
\cite{6}. 
\be
L = \frac{1}{2}(\dot q_2-e^{q_1})^2 + \frac{1}{2}(\dot q_3-q_2)^2\,.
\ee
This model exhibits one primary constraint $\phi^{(0)}=0$, where
\be
\phi^{(0)} = p_1
\ee
The total Hamiltonian ist given by
\be
H_T = e^{q_1}p_2+q_2p_3+\frac{1}{2}p^2_2 + \frac{1}{2}p^2_3 + \lambda p_1\,.
\ee
The model exhibits two secondary constraints which, when generated in the 
way described in section 3, take the form
\be
\phi^{(1)} = -e^{q_1}p_2\,,\quad \phi^{(2)} = e^{q_1}p_3\,.
\ee
These constraints are clearly first class, and their algebra is in strong involution. Following our general procedure 
described in section 3, one is led to the following 
gauge identity:
\be\label{identity5}
\vec v^{(3)}\cdot \vec E^{(3)} - \lambda \vec v^{(2)}\cdot \vec E^{(2)}
+ \lambda^2 \vec v^{(1)}\cdot \vec E^{(1)} \equiv 0\,,
\ee
where the zero modes are of the form (\ref{generic-eigenvectors}), that is,
\bear 
\vec v^{(0)} &=& (\vec 0,\vec 0,1)\,,
\quad \vec v^{(1)} = (-{\tilde\nabla}\phi^{(0)},\nabla\phi^{(0)},0,1)\\
\vec v^{(2)} &=& (-{\tilde\nabla}\phi^{(1)},\nabla\phi^{(1)},0,0,1)\,, 
\quad\vec v^{(3)} = (-{\tilde\nabla}\phi^{(2)},\nabla\phi^{(2)},0,0,0,1)\,,
\eear
where $\vec 0$ is a three-component null vector, 
and the $\vec E^{(\ell)}$ are given by expressions 
of the form (\ref{El}). The gauge identity (\ref{identity5}) can be reduced to 
\ba
v^{(3)}_\alpha E^{(0)}_\alpha &+& 
\frac{d}{dt}\left(v^{(2)}_\alpha E^{(0)}_\alpha\right)
-\lambda v^{(2)}_\alpha E^{(0)}_\alpha 
+\frac{d^2}{dt^2}\left(v^{(1)}_\alpha E^{(0)}_\alpha
 + \frac{d}{dt}(v^{(0)}_\alpha E^{(0)}_\alpha)\right)\nonumber\\
&-&\frac{d}{dt}\left(\lambda v^{(1)}_\alpha E^{(0)}_\alpha
+ \lambda\frac{d}{dt}(v^{(0)}_\alpha E^{(0)}_\alpha)\right)
+ \lambda^2\left( v^{(1)}_\alpha E^{(0)}_\alpha 
+ \frac{d}{dt} (v^{(0)}_\alpha E^{(0)}_\alpha)\right)\nonumber\\
 &&- \lambda\frac{d}{dt}\left( v^{(1)}_\alpha E^{(0)}_\alpha 
+ \frac{d}{dt} (v^{(0)}_\alpha E^{(0)}_\alpha)\right) \equiv 0\,,
\ea
where a summation over $\alpha =1,\cdots,7$ is understood.
Multiplying this identity with $\epsilon(t)$ from the left, and performing the
partial decomposition described in section 3, we may cast this identity into the form (\ref{EdeltaQ}). From there we deduce the transformation laws
\ba
\delta q_i &=& \sum_{\ell=0}^2 \epsilon^{(\ell)}\{q_i,\phi^{(\ell)}\}\,,\cr
\delta p_i &=& \sum_{\ell=0}^2 \epsilon^{(\ell)}\{p_i,\phi^{(\ell)}\}\,,
\ea
where
\ba\label{parametrization}
\epsilon^{(0)} &=& -\ddot\epsilon - 2\lambda\dot\epsilon - \epsilon\dot\lambda - \epsilon\lambda^2\nonumber\,,\\
\epsilon^{(1)} &=& \dot\epsilon + \lambda\epsilon\,,\\
\epsilon^{(2)} &=& -\epsilon\nonumber\,,\\
\delta\lambda &=& \dot\epsilon^{(0)}\,.
\ea
One now readily verifies that the above four parameters satisfy the equations 
(\ref{recursion-relations}) and (\ref{deltalambda}), which in the present case read
\ba\label{recursion}
&&\frac{d \epsilon^{(\ell)}}{dt} + \sum_{\ell'=0}^2 \epsilon^{(\ell')}K^{[\ell'\ell]}=0
\nonumber\,,\\
&&\delta\lambda = \dot\epsilon^{(0)} + \sum_{\ell'=0}^2 \epsilon^{(\ell')}K^{[\ell'0]}=0\,,\nonumber
\ea
where
$ K^{[00]} =  K^{[02]} =  K^{[10]} =  K^{[20]} =  K^{[21]} = 0$,
$ K^{[01]} =  K^{[12]} = 1$ and $K^{[11]} =  K^{[22]} = \lambda$.
This implies in turn, that
$F$ in (\ref{EdeltaQ}) is given by
\be\label{conjecture}
F = -\sum_\ell \epsilon^{(\ell)}\phi^{(\ell)}\,,
\ee
in agreement with our conjecture (\ref{our-conjecture}).
For the variation of the total Lagrangean one finds that $\delta L_T = 0$. 
Note that, in contrast to the example discussed in section 3.2, all the variations
$\delta Q_\alpha$ generate a symmetry of the Hamilton equations of motion,
since they are not proportional to any constraints.
In both examples, however, {\it all} first class constraints play a decisive role for 
generating the symmetries of the total action, and in this sense the 
Dirac conjecture holds for both systems, as expected from our general 
analysis.

\end{document}